\documentclass[twocolumn,showpacs,aps,prb]{revtex4}

\usepackage{graphicx}%
\usepackage{dcolumn}
\usepackage{amsmath}

\makeatletter
\def\btt#1{\texttt{\@backslashchar#1}}%
\DeclareRobustCommand\bblash{\btt{\@backslashchar}}%
\makeatother

\begin{document}

\preprint{}

\title[Short Title]{Metal--Boron Nanotubes}

\author{Alexander Quandt}
 \email{quandt@physics.georgetown.edu}
\author{Amy Y. Liu}
\affiliation{%
Department of Physics, Georgetown University, Washington, DC 20057, USA \\
}%

\author{Ihsan Boustani}
\affiliation{
Institut f{\"u}r Physikalische Chemie, Universit{\"a}t W{\"u}rzburg, 
97074 W{\"u}rzburg, Germany \\
}%


\date{\today}

\begin{abstract}
Nanotubular materials inspired by crystalline diborides
such as AlB$_2$ are proposed. The atomic structure, in particular
the basic chemical question of where to put Al atoms in order 
to stabilize nanotubular Al-B systems, is investigated
using density-functional calculations for prototype
systems.  The optimized tubular prototypes are found to be
competitive in energy with their bulk crystalline counterparts.
All of the tubular Al-B systems investigated are calculated
to be metallic. 
 
\end{abstract}


\maketitle


\section{Introduction}
\label{sec:intro}
Soon after the discovery of carbon nanotubes,\cite{ijima91} it 
was recognized that these compounds can  have electronic properties 
ranging from semiconducting to metallic, 
depending on the details of their atomic structure 
(see chapter 19 in Ref. \onlinecite{dresselhaus96}). 
This has led to much speculation about nanotubular systems 
becoming the materials of choice for future miniaturizations 
of electronic devices towards the nanodomain.\cite{collins00} 
Some major challenges  are the control of the growth mechanism 
in order to obtain carbon 
nanotubes with specific electronic properties,  and the
development of techniques to wire together semiconducting and 
metallic nanotubes in a controlled manner.  One approach for connecting
carbon nanotube devices could be the use of other metallic nanotubes 
that function as wires on the same length scale.   

In an earlier publication,\cite{boustani99}  
we discussed the possible existence of metallic boron 
nanotubes that could serve such a purpose. Boron nanotubes 
can be constructed from the  so-called
{\em Aufbau principle} for the formation of 
stable boron compounds formulated in Ref. \onlinecite{boustani97b}.  This
{\em Aufbau principle} identifies a small set of simple 
structural features characteristic of the most stable isomers 
of pure boron clusters and proposes that highly stable
boron clusters, surfaces, and networks can be constructed using
only these structural elements.  Crystalline $\alpha$-boron  
has long been the paradigm for boron  bonding.  It is comprised of
B$_{12}$ icosahedra with six-fold inverse-umbrella 
coordination,\cite{muetterties67} a configuration that 
obeys the {\em Aufbau principle}.  The 
coordination in the proposed boron  nanotubes is very 
different -- hexagonal pyramidal -- but also follows the
{\em Aufbau principle}.   In a recent study of
clusters of 96 boron  atoms,\cite{boustani00} we showed that the formation  
of large  quasi--planar and tubular systems with six-fold
pyramidal bonding is favored
over the formation of cluster aggregates typical for crystalline 
$\alpha$--boron. We also found that the strain energy
of boron nanotubes relative to the corresponding sheets
is intermediate between those of the carbon and 
boron--nitride systems,\cite{boustani99} 
another favorable condition for the formation of 
boron  nanotubes.
 
In this paper,  the large family of layered  metal diborides
is used as a starting point for investigating
metal-boron nanotubes. 
Diborides of the AlB$_2$ lattice type have graphene-like layers of 
boron atoms, with metal atoms located halfway between the boron layers. 
These structures are similar  to the  pure boron surfaces discussed 
in Ref. \onlinecite{boustani97a}.   In fact, in Ref. \onlinecite{boustani99},
the AlB$_2$ system was used to help elucidate the quasi-planar 
configurations found for pure boron.  Here we extend the  analogy to
the tubular systems.  Density functional calculations are used
to explore the geometry, stability, and electronic structure of
some prototype Al-B nanotube materials.  The prototype nanotubular systems 
presented here are found to be energetically competitive with the 
corresponding bulk counterparts, suggesting that 
synthesis of such nanotubes might be possible. 

The remainder of this paper is organized as follows.
Sec.~\ref{sec:alb2} is devoted to the structural properties 
of the diboride systems. We analyze the AlB$_2$ lattice type,
describe some of the basic physical properties of diborides,
and illustrate the geometric construction of metal--boron 
nanotubes starting from metal--diboride systems. 
In Sec.~\ref{sec:theory}, the technical details of the {\em ab initio}
calculations are given. In Sec.~\ref{sec:results}, 
we present some prototypical Al-B nanotubular structures
that are calculated to be energetically favorable, 
and we discuss the bonding and electronic structure of these
systems.  Conclusions and remarks about future work 
are given in Sec. \ref{sec:conclusions}. The
present study proposes a new class of nanotubular materials
and opens the door to a range of interesting problems for
future studies. 

\section{Construction of nanotubes from crystalline AlB$_2$}
\label{sec:alb2}

\begin{figure}[ht]
\begin{tabular}{c} 
\includegraphics*[angle=-90,scale=0.45]{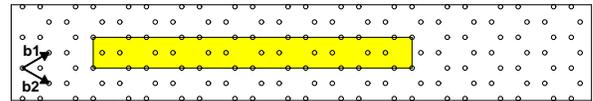} \\
{\bf (a)} Generating a (6,6) carbon nanotube \\ \\ 
\includegraphics*[angle=-90,scale=0.45]{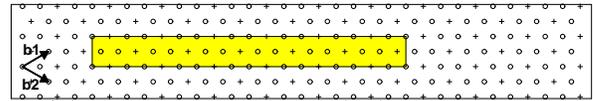} \\
{\bf (b)} Generating a template (6,6) AlB$_2$ nanotube \\
\end{tabular} 
\caption{Cut-and-paste construction for (a) (6,6) carbon nanotubes and  
 (b) template (6,6) AlB$_2$ nanotubes. In (a) circles [$\circ$] denote C
sites; in (b) circles [$\circ$] denote B sites and crosses [$+$]
 denote the projection of Al sites onto the B plane. 
In each case, a (6,6) tube is generated  by rolling up the 
shaded strip, the edges of which are defined by the vector
$6 {\bf b_1} + 6 {\bf b_2}$ and the shortest lattice vector
perpendicular to this direction. 
}
\label{figure1}
\end{figure}

The typical diboride structure is the AlB$_2$ lattice, which 
is composed of two parallel systems of flat layers.\cite{muetterties67} 
One layer contains boron atoms arranged in a honeycomb lattice 
like a graphene sheet (see Fig.\ref{figure1}(a)). 
The Al atoms form a triangular lattice that is located halfway 
between the graphene-like sheets, positioned so that the Al atoms
project down onto the centers of the boron hexagons 
(see Fig.\ref{figure1}(b)).  The space group is $P6/mmm$,  and the
primitive cell is hexagonal with one formula unit per cell. 
Typically, the in-plane lattice constant is around 
$a = 3.1$ \AA.\cite{muetterties67}
This corresponds to an in-plane B-B distance of about 1.8 \AA, 
which is similar to the B-B bondlength in $\alpha$-boron. 
The interplanar B-B distance is significantly larger, with  typical $c/a$
ratios of about 1.1.\cite{muetterties67}

The family of diborides of the AlB$_2$ type comprises materials
with a wide variety of metal atoms substituting for the Al position.  
There are at least 22 members of the family,\cite{muetterties67}
ranging from MnB$_2$ with an effective metallic  
radius of 1.29 {\AA} to GdB$_2$ with an effective metallic radius of 
1.81 {\AA}.  Simple $sp$ elements, as well as $d$- and $f$-electron
atoms can be accommodated.  The ability of the AlB$_2$ lattice 
to incorporate metals that differ so much in size and 
electronic structure is remarkable, and it raises the possibility
of forming alloys between members of this family.

The construction of nanotubular structures from the AlB$_2$ 
lattice type is shown in Fig.~\ref{figure1}. Ignoring the amplitudes of 
the Al atoms for the moment (Fig.\ref{figure1}(b)), 
we see that AlB$_2$ nanotubes can be generated by the same  cut-and-paste 
procedure used to generate carbon nanotubes from graphene 
sheets.\cite{dresselhaus96} 
Nanotubes may be classified by a pair of integers $(N,M)$, 
that specify a cut along a vector $N{\bf b}_1 + M{\bf b}_2$,
where ${\bf b}_1$ and  ${\bf b}_2$ are primitive lattice vectors for
the honeycomb lattice. 
This first cut is followed by 
a second cut perpendicular to this direction that continues 
until another atom of the 
honeycomb lattice is hit. Figure~\ref{figure1} shows the resulting sheet 
for a (6,6) type of nanotube in both systems. Rolling up a general sheet 
and pasting it along the perpendicular direction from both ends of 
the $(N,M)$ cut vector generates the basic unit cell of an $(N,M)$ 
nanotube. 
For the purpose of {\em ab initio} simulations, it is common practice to 
place these tubes side by side on a hexagonal 
superlattice,\cite{dresselhaus96,boustani99} 
thereby approximating the bundles in which these compounds 
form in Nature.

It is interesting to note that if B is substituted for Al in 
Fig.~\ref{figure1}(b), the structure becomes a puckered layer of
six-fold coordinated boron. If 
the construction described above is applied, we obtain the pure
boron nanotubes described in Ref. \onlinecite{boustani99}, where 
we also introduced a different but equivalent construction procedure 
labelled by a pair of integers that describes cuts along the unit 
vectors of a triangular lattice. 
We conjectured that the best way to put
boron nanotubes on a hexagonal superlattice would be to arrange them
such that pairs of boron atoms meet along the directions of the
underlying hexagonal lattice.\cite{boustani99}
Such an arrangement leads to a type of chemical bonding between the tubes
that is analogous to what is found for planar forms of pure 
boron. In the planar systems, pairwise 
bonds between 7-fold coordinated boron sticking out of puckered 6-fold 
coordinated layers of boron were found to be the most favorable arrangement 
for binding single sheets together to form layered systems.\cite{boustani97a}

In earlier studies,\cite{boustani00} we found that tubular boron systems 
could stand alone, {\em i.e.}, like carbon nanotubes, they like to bundle
in the form of a hexagonal superlattice, but they do not have to.
For the diboride system, the situation may be different. 
Studies of small boron clusters  indicate 
that neither the honeycomb lattice nor the related 
carbon--type nanotube structure is stable for boron.\cite{boustani97b} In fact,
the local atomic structure 
of the diboride systems suggests that metal atoms are necessary to 
stabilize honeycomb boron layers or the  tubes derived from such layers.
The metal atoms themselves have to bind to a sufficient number of 
boron atoms (12 in the case of the AlB$_2$ lattice). It is not
clear {\em a priori} whether one could simply bind metal atoms inside or 
outside of a carbon--type boron tube to stabilize
a single tube, or whether those tubes are more likely to come 
in larger bundles of carbon--type boron nanotubes with metal 
atoms both inside and between the tubes, serving in part to bind the
tubes together.  For Al-B tubes, our {\em ab initio} calculations
show that the latter arrangement is preferred, as discussed in 
Sec.\ref{sec:results}. 

\section{Technical Details}
\label{sec:theory}
Calculations were carried out using the VASP 
package,\cite{kresse96a,kresse96b} a
density functional\cite{kohn64,kohn65} based {\em ab initio} 
total-energy code using plane-wave basis sets and ultrasoft 
pseudopotentials.\cite{vanderbilt90,kresse94} 
When performing structural optimizations, 
we let {\em all} degrees of freedom relax, {\em i.e.,} the complete set 
of atomic positions as well as the parameters of the unit cell. The optimal 
configurations as well as the electronic structure for each relaxation 
step were determined using preconditioned conjugate gradient 
procedures.\cite{payne92,teter89} 

The kinetic-energy cutoffs used for the plane-wave expansion 
of electronic wavefunctions were 358.2 eV for C (which we 
use as a  reference system), 161.5 eV for Al, and 321.4 eV for B and all of the 
Al-B systems. The Brillouin zone was sampled on grids of  
(13x13x13) k-points in the cases of crystalline AlB$_2$ and fcc Al,
(8x8x8) k-points in the case of graphite, and (5x5x5) k-points
in the case of all tubular systems with the exception of  the 
(6,6) carbon nanotube for which  (4x4x4) k-points were used. 
These meshes yield meV accuracy in the binding energies, and 
their differing sizes reflect differences in unit-cell sizes.
Care was taken to include a sufficient number of bands above the Fermi 
level to avoid numerical problems in the iterative diagonalization
scheme. 

Although the calculations were well converged with respect to
the number of plane waves and k-points,
the results presented in Sec.~\ref{sec:results} have to be 
interpreted with care.
In particular, the ultrasoft pseudopotential for boron  seems to 
overbind $\alpha$-boron by about 0.7 eV/atom as compared to results 
for the same system 
obtained with harder norm-conserving pseudopotentials.\cite{emin91,lee90} 
On the other hand, 
the cohesive energy we calculate  for the diboride AlB$_2$ 
is in excellent agreement with recent all--electron LDA calculations 
on diboride systems.\cite{ivanovskii00} This suggests that the 
ultrasoft pseudopotential for boron has 
difficulty describing the inverse-umbrella bonding in $\alpha$-boron, 
but it nevertheless gives an accurate description for nearly planar 
configurations of boron like the ones found in the diborides, 
quasi-planar boron sheets, and their related nanotubes. 

Another potential problem is the fact that the conjugate gradient 
optimization does not guarantee the detection of a global minimum.
Given the reported pathology of the energy hypersurfaces of boron 
clusters,\cite{boustani97b} it is likely  that for the large systems 
discussed below, we relax into local minima rather than 
global minima. However, as discussed in Ref. \onlinecite{boustani97b}, 
it is also likely that the relaxed structures capture the right 
chemistry, and all we have to 
worry about is the fact that the cohesive energies corresponding to these local 
minima will be slightly higher than those of the global minimum.   

\section{Results and discussion}
\label{sec:results} 

Unlike pure boron and pure carbon nanotubes, the Al-B nanotubes we
investigated do not appear to be stable as stand-alone tubes.
We were unable to converge structural optimizations  
of isolated tubes constructed
by rolling up puckered sheets of AlB$_2$.  Furthermore,
in structural optimizations of Al-B nanotubes on periodic
superlattices, the tubes always ended up  packed closely together, 
even if they started out well separated.  
Hence, for the remainder  of this paper, we  focus
on results for Al-B tubes arranged on a hexagonal superlattice.  
For comparison, parallel calculations were
performed for pure boron and pure carbon nanotubes.  In each case, the (6,6) 
nanotube was used as a prototype. 

The first step was to generate reference data for atomic structure 
and cohesive energies. This data is listed in the lower part of 
Table~\ref{table1}. Fcc Al and $\alpha$-boron serve as
basic reference structures for the Al-B systems.
Phase-separated 
AlB$_2$ (into fcc Al and $\alpha$-boron) is calculated to have a  
cohesive energy of 6.38 eV/atom. As can be seen in Table~\ref{table1},
crystalline AlB$_2$ is calculated to be only slightly stable against
phase separation.   However, the validity of this result is questionable
in light of the known stability of crystalline AlB$_2$.\cite{muetterties67}  
As we did a full relaxation on all systems, with high plane-wave cutoffs 
and many k-points, we attribute  the problem to the 
overbinding of $\alpha$-boron discussed in Sec. \ref{sec:theory}. 
As already mentioned, the cohesive energy of AlB$_2$ obtained in a 
recent all-electron calculation\cite{ivanovskii00} is in excellent 
agreement with our value listed in Table~\ref{table1}. 

The same considerations apply when we examine the stability of the
(6,6) boron nanotube.  The results in Table~\ref{table1} indicate 
that the (6,6) boron nanotube is 
unstable by 0.5 eV/atom compared to crystalline $\alpha$-boron.
This is in contrast to our recent all--electron {\em ab initio} 
calculations, which found that for B$_{96}$ clusters, the 
nanotubular isomers challenge or exceed the stability of cluster aggregates 
found in $\alpha$-boron.\cite{boustani00} 
Again, we attribute this discrepancy to the overbinding
of $\alpha$-boron  by the ultrasoft pseudopotential used.

\begin{figure*}[h]
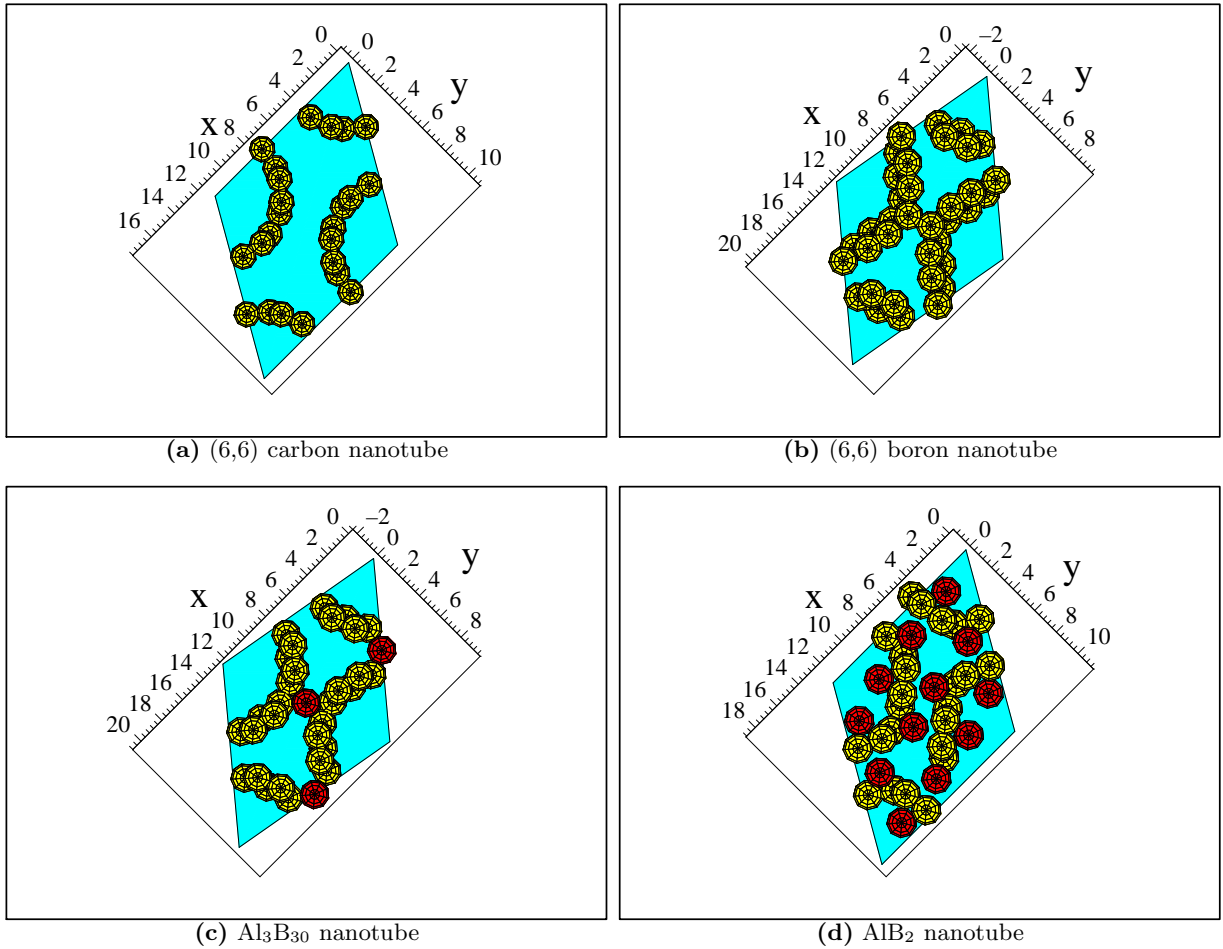

\begin{tabular}{cc} 
\includegraphics*[angle=-90,scale=0.35]{carbon66.epsi} & 
\includegraphics*[angle=-90,scale=0.35]{bortub.epsi}\\
{\bf (a)} (6,6) carbon nanotube & {\bf (b)} (6,6) boron nanotube  \\ 
\includegraphics*[angle=-90,scale=0.35]{al3b30tub.epsi} & 
\includegraphics*[angle=-90,scale=0.35]{alb2tub.epsi} \\ 
{\bf (c)} Al$_3$B$_{30}$ nanotube & {\bf (d)} AlB$_2$ nanotube \\
\end{tabular} 
\caption{Comparison of various optimized tubular structures. 
Each panel shows one 
unit cell of the hexagonal superlattice on which tubes are arranged. 
In  (c) and (d), the light atoms are B and  
the dark atoms are Al. Units of the surrounding boxes are given in \AA.  
Note that even after relaxation, the atoms remain stacked in two layers
only.}
\label{figure2}
\end{figure*}

The situation looks more promising when we analyze the 
{\em chemistry} of the boron nanotubes.  The relaxed structure for
the (6,6) boron nanotube is shown in Fig.~\ref{figure2}(b). 
Note that the representation of the tubes in Fig.~\ref{figure2} is 
perhaps unusual in that no tube is shown in its entirety. 
Instead, this representation is used to emphasize 
the chemical bonding between nanotubes sitting on the hexagonal superlattice. 
Prominent boron to boron bonding along the direction of the lattice
is evident in Fig.~\ref{figure2}(b).  Further, there is
a slight faceting due to undulations from the basic hexagonal lattice of 
the quasiplanar reference structure. Both of these features
are in perfect agreement with the structure and the chemical 
bonding found in boron sheets.\cite{boustani97a} 
Therefore, it is reasonable to take the relaxed (6,6) boron nanotube 
as a good starting structure when looking for nanotubes of the Al-B type. 

Table~\ref{table1} includes an additional set of benchmarks. By comparing 
graphite and the (6,6) carbon nanotube (Fig.~\ref{figure2}(a)), we get a 
realistic measure for the energetic difference between comparable carbon 
nanotubes and their layered reference structure. We calculate 
the (6,6) carbon nanotube to be about 0.14 eV/atom higher
in energy than graphite.  Tubes with larger radii would have even 
smaller instabilities.\cite{dresselhaus96}  

Among the many variants of Al-B nanotubes that we explored, we focus 
on the two shown in Figs.~\ref{figure2}(c) and (d).  
Our approach was to start with the relaxed B nanotube and substitute
Al for B.  The structure shown in Fig.~\ref{figure2}(c) 
has composition Al$_3$B$_{30}$ and was generated from the (6,6) nanotube 
by removing pairs of boron atoms located along the directions of the 
hexagonal lattice, and substituting them by single Al atoms. 
This can happen at three different places within the unit cell, and the 
Al atom then sees a local atomic environment similar to AlB$_2$. 
Indeed, after full relaxation, the system tends to recreate these 
local environments, and the tube as a whole becomes less puckered 
(compare Fig.~\ref{figure2}(c) to  the puckered B tube in
Fig.~\ref{figure2}(b) and the unpuckered C tube in Fig.~\ref{figure2}(a)).
However, the energetic gain is not substantial: the cohesive
energy of phase-separated Al$_3$B$_{30}$ is about 7.2 eV/atom, 
so there is only a small improvement in stability as compared to the 
(6,6) boron tube. 

Figure~\ref{figure2}(d) shows an Al-B nanotube system with
stoichiometry AlB$_2$.  This configuration is calculated to be 
unstable by only 0.23 eV/atom compared to crystalline AlB$_2$. 
While this energy difference is not quite as  small as
that  between the (6,6) carbon tube and graphite, it is close. 
The structure was generated from the (6,6) boron nanotube
by substituting Al for boron at the six positions located along the 
directions of the hexagonal superlattice, which are on the
outside layer of the puckered tube, and at six positions
on the inside layer of the tube.  In the optimized structure,
the tubes are rotated from the orientations they adopt in the
superlattice of pure boron nanotubes in such a way that Al 
atoms sticking out from 
adjacent tubes do not face each other, but rather sit
between boron layers.  Such an arrangement
locks the tubes together in a gear-like fashion, and 
creates a local environment similar to that in crystalline AlB$_2$.
The inclusion of Al atoms inside the tube is found to be 
important for stabilizing this structure. 

\begin{figure*}[h]
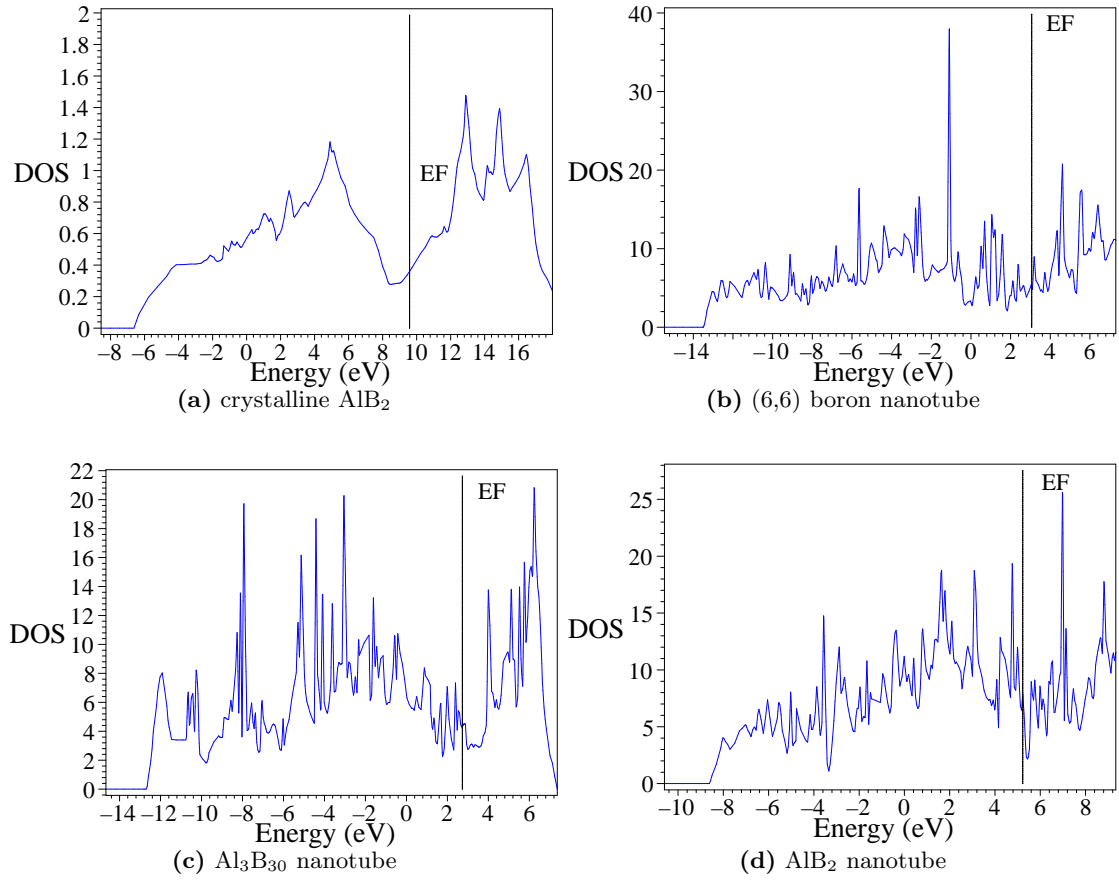

\begin{tabular}{cc} 
\includegraphics*[angle=-90,scale=0.35]{alb2dos.epsi} & 
\includegraphics*[angle=-90,scale=0.35]{bortubdos.epsi} \\
{\bf (a)} crystalline AlB$_2$ & {\bf (b)} (6,6) boron nanotube  \\ 
& \\
\includegraphics*[angle=-90,scale=0.35]{al3b30tubdos.epsi} & 
\includegraphics*[angle=-90,scale=0.35]{alb2tubdos.epsi} \\ 
{\bf (c)} Al$_3$B$_{30}$ nanotube & {\bf (d)} AlB$_2$ nanotube \\
\end{tabular} 
\caption{Densities of state (DOS) in states/eV/cell vs. energy in eV for 
crystalline AlB$_2$ and various optimized tubular systems. All systems 
are metallic.  Similarities in the DOS arise from 
similarities in  local atomic environments.}
\label{figure3}
\end{figure*}

The electronic densities of states  (DOS) calculated for crystalline
AlB$_2$ as well as the (6,6) B, Al$_3$B$_{30}$, and AlB$_2$ tubes
are plotted in Fig.~\ref{figure3}.  All of the tubes are metallic. 
This result was predicted in Ref. \onlinecite{boustani99} based on 
qualitative arguments related to hexagonal tight-binding 
models,\cite{pettifor95} as well as on static 
{\em ab initio} calculations. The present results show that
these predictions even hold after a complete structural
relaxation. 
The overall shape of the DOS for crystalline
and tubular AlB$_2$ are similar due to similarities in  local atomic 
environments.  The DOS of the AlB$_2$ nanotube, however,  has 
much more fine structure, which can be understood in terms
of a backfolding of the bands of crystalline AlB$_2$ into the 
reduced Brillouin zone of the tubular lattice. 
The DOS of Al$_3$B$_{30}$  plotted in Fig.~\ref{figure3}(c) on the other 
hand shows some similarity to the DOS of the (6,6) boron nanotube depicted 
in Fig.~\ref{figure3}(b), with some distortion due to the modest content of 
Al atoms.

\section{Conclusions}
\label{sec:conclusions} 
We have explored the plausibility of metal-boron nanotubular systems 
derived from diboride materials. Our data suggests that it
may be possible to synthesize such systems. We predict that 
metal--boron nanotubular materials will likely form in
bundles built from rolled up graphene-like sheets of boron, 
with metal atoms sitting inside and outside the tubes, 
preserving as much as possible the local atomic environments 
found in the AlB$_2$ lattice.  {\em Ab initio} calculations 
of their basic electronic properties show that such tubular metal-boron 
systems are metallic.  

There are numerous issues of interest for future studies. 
First, a comparison of the properties expected 
for nanotubes derived from different diboride compounds -- including
an assessment of which, if any, form stand-alone tubes --  would
be of interest.  Continuing along those lines, it may be possible
to tune some of the properties of the metal-boron tubes by forming
alloys.  Alternatively, substitution of boron atoms 
by carbon is likely to have a serious impact on the electronic structure 
of the resulting tubular materials, bringing them closer to 
BN materials.\cite{muetterties67}  Also, the recent 
discovery of superconductivity in MgB$_2$ raises the question
of whether some of the  diboride-derived tubular materials
might superconduct.\cite{akimitsu01}  
Finally, and most importantly, we look forward to 
cooperating with experimental groups that may be able to
synthesize tubular metal-boron materials.

\begin{acknowledgments}
We thank Rodney Ruoff for valuable  discussions
and for his encouragement to follow up on our study of the
diboride systems. 
This work was supported by the National Science Foundation under Grant
DMR-9973225.   AYL acknowledges support from the 
U.S. Navy -- ASEE Faculty Sabbatical Leave Program. 
\end{acknowledgments} 


\begin{references}

\bibitem{ijima91}
S. Ijima, Nature {\bf 354}, 56 (1991). 

\bibitem{dresselhaus96}
M. S. Dresselhaus, G. Dresselhaus, and P. C. Eklund,
{\em Science of Fullerenes and Carbon Nanotubes},
(Academic Press, London, 1996).

\bibitem{collins00}
P. B. Collins and P. Avouris, 
Sci. Am. {\bf 283}, 62 (2000). 


\bibitem{boustani99}
I. Boustani, A. Quandt, E. Hernandez, and A. Rubio,
J. Chem. Phys. {\bf 110}, 3176 (1999).

\bibitem{boustani97b}
I. Boustani, Phys. Rev. B {\bf 55}, 16426 (1997).


\bibitem{muetterties67}
 E. L. Muetterties, {\em The Chemistry of Boron and its Compounds},
(Wiley, New York, 1967).

\bibitem{boustani00}
I. Boustani, A. Quandt, and A. Rubio, J. Solid State Chem. {\bf 154},
269 (2000).


\bibitem{boustani97a}
I. Boustani, Surf. Sci. {\bf 370}, 355 (1997). 


\bibitem{kresse96a}
G. Kresse and J. Furthm{\"u}ller, Comput. Mater. Sci. {\bf 6}, 15 (1996).

\bibitem{kresse96b}
G. Kresse and J. Furthm{\"u}ller, Phys. Rev. B {\bf 54}, 11169 (1996).

\bibitem{kohn64}
P. Hohenberg and W. Kohn, Phys. Rev. {\bf 136}, 864 (1964).

\bibitem{kohn65}
W. Kohn and L. J. Sham, Phys. Rev. {\bf 140}, 1133 (1965).

\bibitem{vanderbilt90}
D. Vanderbilt, Phys. Rev. B {\bf 41}, 7892 (1990). 

\bibitem{kresse94}
G. Kresse and J. Hafner, J. Phys.: Condens. Matter {\bf 6}, 8245 (1994).


\bibitem{payne92}
M. C. Payne, M. P. Teter, D. C. Allan, T. Arias, and J. D.
Joannopoulos, Rev. Mod. Phys. {\bf 64}, 1045 (1992).

\bibitem{teter89}
M. P. Teter, M. C. Payne, and D. C. Allan, Phys. Rev. B {\bf 40},
12225 (1989). 

\bibitem{emin91}
L. Kleinman, in {\em Boron Rich Solids}, edited by 
D. Emin, T. L. Aselage, A. C. Switendick, B. Morosin, and C. L. Beckel
(American Institute of Physics, New York, 1991),
vol.  231 of {\em AIP Conference Proceedings}. 

\bibitem{lee90}
S. Lee, D. M. Bylander, and L. Kleinman, Phys. Rev. B {\bf 42}, 1316 (1990).

\bibitem{ivanovskii00}
A. L. Ivanovskii and N. I. Medvedeva, Russian J. of Inorganic Chem. {\bf 45},
1234 (2000).

\bibitem{pettifor95}
D. Pettifor, {\em Bonding and Structure of Molecules and Solids}
(Oxford University Press, Oxford, 1995). 


\bibitem{akimitsu01} 
J. Magamatsu, N. Nakagawa, T. Muranaka, Y. Zenitani, and J. Akimitsu, 
Nature {\bf 410}, 63 (2001). 

\end{references}

\vspace*{1cm}
\newpage

\begin{table}[t]
\caption{Number of atoms per unit cell $n$, and cohesive 
energies $E_{coh}$, for pure and mixed boron nanotubes and several 
reference structures. 
\label{table1}}
\begin{tabular}{lcc} 
\colrule
\colrule
System  & $n$ & ~~~~ $E_{coh}$ [eV/atom] ~~~~\\  
\colrule
(6,6) boron nanotube~~~~~~~~~~ & 36 &     7.001 \\ 
Al$_3$B$_{30}$ nanotube & 33 &  6.884 \\ 
AlB$_2$ nanotube & 36 &         6.227 \\ 
\colrule 
$\alpha$-boron & 12 &           7.509 \\ 
fcc Al  & 1 &                   4.119  \\ 
AlB$_2$ & 3 &                   6.451 \\ 
\colrule
graphite & 4 &                 10.156 \\ 
(6,6) carbon nanotube & 24 &   10.019 \\  
\colrule
\colrule
\end{tabular}
\end{table}


\end{document}